\documentstyle[epsf,12pt]{article}
\begin{document}
%\draft
\newcommand \pt {{\tilde p}}
\newcommand \bQ {{\bf Q}}
%\twocolumn[\hsize\textwidth\columnwidth\hsize\csname@twocolumnfalse\endcsname

\title{First steps of a nucleation theory in disordered systems}
\author{Silvio Franz \\
 \\
 {\it  The Abdus Salam International Center for Theoretical Physics}\\
{\it Strada Costiera 11, P.O. Box 563,
I-34100 Trieste (Italy)}\\
{\it E-mail: franz@ictp.trieste.it}}

\date{\today}

\maketitle

\vspace{.5 cm}
\begin{abstract}
We devise a field theoretical formalism for a microscopic theory of
nucleation processes and phase coexistence in finite dimensional glassy
systems. We study disordered $p$-spin models with large but finite range of
interaction. We work in the framework of glassy effective potential theory
which in mean-field is a non-convex, two minima function of the overlap. We
associate metastability and phase coexistence with the existence of space
inhomogeneous solutions of suitable field equations and we study the simplest
of such solutions.
\end{abstract}
%\end{abstract}
%\pacs{05.20, 75.10N}

%\twocolumn\vskip.5pc]\narrowtext

\section{Introduction}
A prominent theoretical problem in the physics of glassy systems is the
comprehension of the processes that restore ergodicity in regions where
approximated liquid theories or
Mean-Field models predict a spurious structural arrest.  It is well known
that the Mode Coupling Theory \cite{MCT}, while describing correctly several
aspects of the dynamics of moderately supercooled liquids, predicts the
divergence of the relaxation time and ergodicity breaking at a temperature
$T_c$ well above the observed laboratory glass transition temperature
$T_G$. In the so called ``random first order phase transition scenario''
\cite{W1}, corresponding to ``one step replica symmetry breaking'' (1RSB)
\cite{mpv} in mean field disordered models, this ergodicity breaking appears
to be related to the appearance of exponentially numerous metastable states,
capable to trap the system in regions of free-energy extensively higher than
the thermodynamic value. Within this scenario a Kauzmann-like entropy crisis
occurs at a temperature $T_K<T_c$.  Below this temperature, an ideal glassy
state is thermodynamically stable. 

Of course, in extended systems, strict metastability is an artifact of
mean-field approximation and should disappear as soon as the finite range
character of interaction is properly treated. A remarkable hypothesis states
that while acquiring a finite life time, the gross structure of metastable
states survives in short range glassy systems and dominates the low
temperature relaxation.  Unfortunately, a microscopic satisfactory description
of the configuration space for liquid systems below $T_c$ is presently
lacking. Based on the random first order transition scenario, Kirkpatrick,
Thirumalai, Wolynes and others \cite{W1,W2,W3,W4,W5} have proposed the notion
of ``mosaic state'' where below a well defined temperature-dependent
coherence length the system appears to behave in a mean-field like glassy
manner, while it behaves as a liquid on larger scales. This length should be
divergent at the temperature $T_K$, where the ideal glassy phase sets in. We
meet another much debated problem in the theory of disordered systems: the
possibility in finite dimension of having ideal glassy phases with
characteristics similar to the mean field ones \cite{marinari,ns,moore}.

In the past, different attempts have been made to investigate the
restoration of ergodicity in the temperature region between $T_K$ and $T_c$
and the description of the low temperature phase $T<T_K$, through analogies
with first order phase transitions. In the aforementioned contributions
\cite{W1,W2,W3,W4,W5}  a phenomenological nucleation theory was proposed,
where the exponential multiplicity of metastable states would give rise to
``entropic droplets'' providing the main driving force for ergodicity
restoration. This force would be contrasted by an interface cost proportional
to a power $d-d_c$ of the droplet radius. In the simplest version of the
argument, $d_c$, which can interpreted as the lower critical dimension for an
ideal glassy phase, is equal to one and the resulting free-energy barrier
scales as $(T-T_K)^{-2}$ in three dimensions. A great effort was made to
justify a renormalization of the value of $d_c$, in order to derive a
Vogel-Fulcher scaling $(T-T_K)^{-1}$.

It was later realized by Parisi \cite{P1,P2} that a nucleation theory in
glassy systems could be formalized considering an appropriate effective
potential function \cite{FPV1,FP1,FP2,FP3} exhibiting a characteristic double
well shape below $T_c$.  Postulating a finite surface tension between
metastable states, in \cite{P1,P2} the mean field behavior of the critical
droplet in the form anticipated in \cite{W1,W2} was derived. The first
trials to compute from first principle entropic droplets were performed in
\cite{W6} for a random heteropolymer model.  

Biroli and Bouchaud \cite{BB}
have recently rephrased some of the concepts in \cite{W1,W2,W3,W4,W5},
elucidating at the physical level the notions of mosaic state and entropic
droplets, and proposing a self-consistent derivation of a glassy coherence
length.  Again, an interface cost to put different metastable states in contact
is supposed rather then derived.

In ordinary first order phase transitions with two or more phases in
competition, metastable states are destabilized by nucleation
processes.  These are possible as soon as the interaction range is
finite. A proper theory of nucleation is achieved considering a large
but finite interaction range, where appropriate asymptotic expansions
can be applied \cite{L1,L3,L4}.  The study of interfaces and
nucleation is reduced to the analysis of space-inhomogeneous
solutions of the saddle point equations and fluctuations around these
solutions. This method can be seen as an expansion around mean-field
and becomes more and more accurate for larger and larger interaction
range.

The aim of this paper is to set up a formalism allowing for a first principle
ergodicity restoration and phase coexistence theory in disordered systems,
starting from the theoretical analysis of microscopic models.  The goal of
this theory should be the computation of free-energy barriers and the critical
dimension $d_c$. The crucial idea, in common with the previous
phenomenological analysis, is that in disordered systems local properties can
behave similarly to mean-field, while this is not necessarily so for global
properties. This idea has recently received a rigorous foundation in the
context of spin glass models with large but finite range Kac kind of
interactions \cite{FT1,FT2,FT3}.  The analysis of these models, provides
therefore a natural starting point to study finite range effects in expansions
around mean field. In this paper, we use long but finite range Kac versions of
the spherical $p$-spin models and analyze them by the means of the replica
theory. The general framework in which we move is the setting of effective
potential theory for coupled replica systems \cite{FPV1,FP1,FP2,FP3}.  This
allows to define field theoretical free-energies functionals of local
overlaps, analogous to the familiar Landau free-energies functional of the
local-order parameter in non disordered systems. In the presence of a ``random
first order transition'' the potentials have a two minima structure similarly
to systems undergoing a first order phase transition.  We argue that one can
devise a theory of ergodicity restoration and phase coexistence which,
analogously to nucleation in first order transition, is based on the analysis
of the instantoinc, space inhomogeneous solutions of suitable field equations.
In this paper we begin the analysis of these equations both in the metastable
and in the coexistence region. This allows us to put on formal basis the
extensions to disordered systems of the concepts of critical droplets,
interface tension and associated free-energy barrier.  In spite of the fact
that the content of the theory we develop is very different from ordinary
nucleation of a stable phase into an metastable one, thanks to the many formal
analogies, we found it useful to use the intuitive language of first order
transitions. To avoid confusion we always specify the physical meaning of the
quantities involved. Due to the complexity of the resulting theory, we expect
many possible solutions to exist to the instantonic equations.  In this paper
we always look for the simplest of such solutions. It is well possible
that in the future better solutions will improve our results. In addition,
while a complete theory should include the study of the matrix of small
fluctuations around the instantonic saddle points, we do not perform this
analysis here leaving it for future research.\footnote{After this paper
appeared as a preprint (cond-mat/0412383) an interesting contribution by
M. Dzero, J. Schmalian, P.  G. Wolynes (cond-mat/0502011) proposed a partial
analysis of the fluctuations matrix, as well as a new solution in one of the
cases considered here.}

This paper comprises 7 sections. In the second section we briefly
review the effective potential method for mean-field systems. In
section 3 we introduce the model. In section 4 we discuss the
free-energy functional in the replica approach and in section 5 we
discuss our results. Finally we summarize our conclusions. Some
technical aspects of our analysis are discussed in an 
appendix. 

\section{The effective potential theory}

As customary in statistical physics \cite{Zinn,parisi-book} we 
consider coarse grained free-energy functionals as functions of the
local order parameter as a starting point in the study of medium and
large distance properties of our system.  In mean-field disordered
glassy systems the order parameter is the probability distribution of
the overlap between configurations, as induced by the canonical
measure and the quenched disorder \cite{mpv}. In spin systems the
overlap is just the normalized scalar product among spin
configurations, while other notions of similarity have been proposed
for particle systems \cite{parisi-ove}.  Proper effective potential
functions for glassy systems have been obtained considering the
free-energy of identical copies of a given system for fixed values of
the mutual overlaps.  The constraint on the overlaps allows for
suitable modifications of the Boltzmann weight so as to access
otherwise hidden regions of configuration space. The method has been
successfully used to study metastable states in mean field spin glass
models, $p$-spin models \cite{FP1,FP2,FP3} and ROM \cite{ritort}, and
liquids in the HNC approximation \cite{CFP1,CFP2} with similar
results. Models undergoing 1RSB and in particular $p$-spin models, are
the prototype of mean-field systems displaying a Kauzmann-like entropy
crisis.  The configurational entropy, defined as the logarithmic
number of ergodic components that contribute at each temperature to
the partition function, is an increasing function of the
temperature in the domain $T_K<T\leq T_c$ and vanishes at $T_K$.

This paper  considers local versions of effective
potentials for spatially extended systems.  Before entering into the
discussion of these versions, we briefly review the general
properties of the potentials in mean field. Full details can be
found in \cite{FP1,FP2,FP3}.

Two main versions are usually considered: the {\it annealed potential}
and the {\it quenched potential } that allow to weigh configuration
space regions in different ways and thus study different aspects of
the glassy phase.

In the annealed version one considers $r$ copies of the original
system and study the constrained free-energy where only the
configurations such that all the mutual overlaps have a fixed value
$\pt$ are taken into account. In this paper we limit ourselves to
considering the most commonly studied case $r=2$, which, denoting
$q(\sigma,\tau)$ the overlap between configurations $\sigma$ and
$\tau$, can be written with transparent notation as
\begin{eqnarray}
V_a(\pt)=-\frac{T}{N}E \log\left[\frac{1}{Z^2}
\sum_{\sigma,\tau} \exp\left(-\beta
  (H(\sigma)+H(\tau))\right) \delta\left(q(\sigma,\tau)-\pt\right)
\right].
\end{eqnarray}
In words, this is the average $E$ over the quenched disorder of the 
logarithm of the probability distribution of the overlap  among two replicas
\cite{mpv}. 

In the quenched construction one fixes an unconstrained equilibrium
configuration and considers the free-energy of a second system which
has a fixed overlap $\pt$ with this reference state.
This is written as 
\begin{eqnarray}
V_q(\pt)=-\frac{T}{N}E \frac{1}{Z}\sum_\tau \exp \left(-\beta
  H(\tau)\right)  \log\left[\frac{1}{Z}
\sum_{\sigma} \exp\left(-\beta
  H(\sigma)\right) \delta\left(q(\sigma,\tau)-\pt\right)
\right]. 
\end{eqnarray}

\begin{figure}
\begin{center}
\epsfxsize=250pt
\epsffile{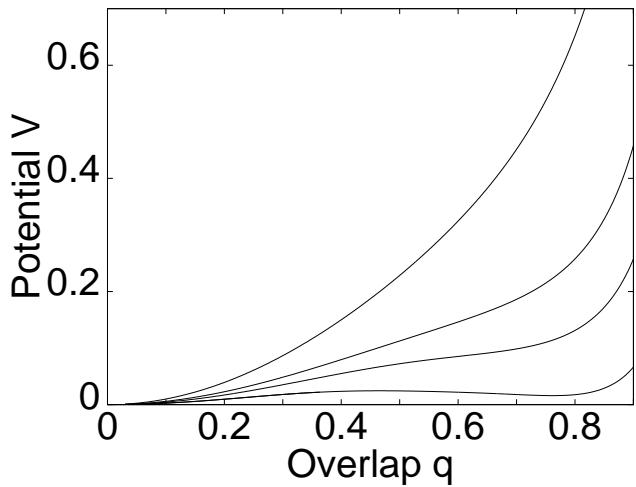} 
\end{center}
  \caption[0]{\protect\label{pot}  
Qualitative behavior of the     
quenched potential for the spherical $p$-spin at various
    temperatures (see the text). }
\end{figure}

In both cases the potential exhibits the same qualitative features,
analogous to the ones of the mean field free-energy as a function of
the order parameter in presence of a first order phase transition.  At
high temperature it is a convex function with a single minimum for
zero value of the overlap. On decreasing the temperature it first
looses convexity and then at a temperature $T_s$ which depends on the
version one is considering, it develops a secondary minimum at
higher values of the overlap.  Finally, at
the even lower temperature $T_K$ of the aforementioned entropy crisis,  
and below, both potentials exhibit two minima with
degenerate free-energies (see fig. \ref{pot}).

Differently from ordinary first order phase transitions however, the two minima
are not associated to phases with different macroscopic characteristics but
rather, the whole shape of the function is related to the existence of an
exponential multiplicity of ergodic components. In particular, the
difference in free-energy between the secondary and the primary minimum
represents different quantities for the annealed and the quenched potential,
but in both cases it is associated with the existence of a multiplicity of
metastable states with different internal free-energy.

The quenched potential as a function of the temperature strictly reflects the
phase structure of the model \cite{FP1}. The temperature $T_s$ where the
secondary minimum first appears coincides here with the ergodicity breaking
temperature $T_c$ \cite{cris-som}.  It is well known that below $T_c$ the
equilibrium free-energy can be decomposed in the sum of an internal
free-energy, i.e. the free-energy of single ergodic components (which is equal
for all the components relevant at a given temperature), and the
configurational entropy, the log-multiplicity of these metastable states
\cite{Abarrat}.  While the internal free-energy calculated in the two minima
is the same, the difference in absolute free-energy directly measures the
configurational entropy multiplied by the temperature. At the temperature
$T_K$ where the configurational entropy vanishes, the two minima become
degenerate, and remain degenerate below.

In the $r=2$ annealed potential, the secondary minimum appears at a
temperature $T_s=T_2>T_c$. It was found in \cite{FP3} that the secondary
minimum is associated to both copies being in one of the lowest internal
free-energy states: the ones of zero configurational entropy. In this case the
difference between the secondary and the primary minima represents the cost
needed to put a system in the lowest available internal free-energy
states. This is a positive cost, since this situation implies a loss in
configurational entropy.  As in the quenched case, for temperatures equal to
$T_K$ and below, the two minima are degenerate. It should be noticed that
close to $T_K$, while in the quenched case one has that the free-energy
difference among the minima behaves as $T-T_K$, in the annealed case it
behaves as $(T-T_K)^2$. It is less costly to constrain a system in the
internal free-energy ground state then in a single particular equilibrium
state. At and below $T_K$ the minima of the two potentials give the same
information. In both cases they probe the lowest free-energy components of the
system.  Due to this fact, we feel it
important to discuss the two cases in parallel.

Both versions can be adapted to spatially extended systems to define
free-energies as functionals of local order parameters. Once the
free-energy functional is built, one can use it to study
metastability, ergodicity restoration and phase coexistence.

An intuitive argument on the existence of a coherence glassy length has been
put forward in \cite{BB}, that can be directly rephrased in terms of the
quenched potential. If one wishes to know over which length scale a
supercooled liquid appears to be instantaneously frozen, one can consider the
configurations of a system constrained to be ``close'' to a reference
equilibrium state outside a bubble of radius $R$. For low $R$ it is
free-energetically advantageous to remain close to the reference state while
for large $R$ it is advantageous to get far apart from the
reference. Biroli and Bouchaud identify the coherence length as the radius
$R^*_q$ that separates the two regions. This radius can be computed as the
critical radius for an overlap droplet in the quenched setting. Certainly, one
should specify what ``close'' means in this context. The natural choice for
the Kac model is to chose it to be at the overlap $\pt^*$ where the
potential has the secondary minimum.

Obviously the same line of reasoning can be applied to the annealed
case. How large can  a bubble be such that two replicas that are
constrained to be at overlap $\pt^*$ outside the bubble, will also be
close inside the bubble? This problem is related to the existence of a
phase transition at $T_K$. In case of a transition one would expect a
divergent critical radius, and a divergent length (long range order)
in the whole low temperature phase $T<T_K$. This divergent length is
usually associated with a divergent free-energy barrier.  One gets then
an estimate of the lower critical dimension as the highest dimension
where this barrier fails to diverge. From the above discussion we see that in
the effective potential setting the problem can be formally expressed as
 a nucleation theory where one seeks  instantonic
configurations of the overlap in space. Unlike ordinary nucleation, 
the droplet size should not be interpreted as the size of the typical
fluctuation inducing the decay of a metastable phase, but rather as the 
typical length over which there is a crossover from glassy-like to liquid-like
behavior \cite{W1}.

\subsection{Technical matter} 

In disordered systems, the annealed and quenched potentials can be
studied using the replica method.  One can deal with both cases in a
unified framework considering a number $n'=nr$ replicas, with $n$
going to zero in the usual way.

 In
the annealed case $r$ is considered to be a fixed integer number,
while $n$ tends to zero. The number of real copies $r$ is usually
taken to be equal to 2, and this is the case we discuss in this paper.

In the quenched case one gets the effective potential from an
additional analytic continuation in $r$.  What one needs to do, is to
take the $r$ derivative of the replicated free-energy in the point
$r=1$ \cite{FP1}.  

In mean-field the order parameter in the replica formalism is the
$n'\times n'$ overlap matrix $\bQ_{\alpha,\beta}$.  The free-energy as
a function of $\bQ$ admits an expression formally identical to the one
of the unconstrained problem \cite{FP1}. The difference with the unconstrained
problem lies in the fact that some of the elements of the matrix
$\bQ$ are fixed to the value $\pt$ of the constrained overlap. It is
useful to view $\bQ$ as a collection of $r^2$ $n\times n$ sub-matrices
$Q_{a,b}^{u,v}$ with $u,v=1,...,r$ and $a,b=1,...,n$. The overlap
constraint in the partition function is reflected in the fact that the
elements $Q^{1,u}_{a,a}$ and $Q^{u,1}_{a,a}$ for all $u$ and $a$ are
fixed to the value $\pt$.  The spherical constraint implies
$Q_{a,a}^{u,u}=1$ for all $u$ and $a$.  The analytic continuation for
$n\to 0$ can be performed supposing Parisi Ansatz for each of the
sub-matrices $Q^{u,v}=Q^{v,u}$ and the following ``replica symmetric
structure'' in the upper indexes\footnote{Since the replicas with
  $u=1$ are singled out due to the constraint, a matrix respecting
  replica symmetry in the upper indexes is only invariant under
  permutation of the indexes $u=2,...,r$.}: $Q^{1,1}=Q$, $Q^{1,u}=P$
and $Q^{u,u}=T$ for $s>1$ and $Q^{u,v}=S$ for $u\ne v$ and $u,v>1$.
Replica symmetry breaking in the upper indexes could also be possible,
but will not be discussed in this paper. Our choice is a valid Ansatz
for all $r$ and allows the continuation $r\to 1$.  In the annealed
case $r=2$, the matrix $S$ is obviously absent, and by symmetry one
has $T=Q$.

In the following we mainly use the one step replica symmetry
breaking form for the $n\times n$ replica matrices, where a given matrix
$A_{ab}$ has a block structure with $n/m$ blocks of size $m$
and is parameterized by the common value ${\tilde a}$ of its
diagonal elements, its value $a_1$ inside the blocks and its value $a_0$ 
outside the blocks \cite{mpv}. It is customary to represent such a
form with a function $a(u)=a_0 \theta (m-u)+a_1\theta(u-m)$ where $u$
is a variable in the interval $[0,1]$.

\section{The model}

In order to introduce physical space in the theory, 
we consider in this paper a finite-dimensional version of
the spherical p-spin model \cite{cris-som} on a
$d$-dimensional cubic lattice $\Lambda$ of linear size $L$, with spin
variables $S_i$ defined on each lattice site $i\in \Lambda$. We use a
spatially local spherical constraint. To this aim we partition
$\Lambda$ into cubes of size $l$ and on each cube $B_n$
($n=1,...,(L/l)^d$) we impose the constraint: $\sum_{i\in B_n}
S^2_i=l^d$. The reason for this construction is that, as for the
global spherical constraint in mean-field, it will allow to write the
free-energy as a function of the order parameter matrix in a closed
form. We then introduce the finite range $p$-spin Hamiltonian \cite{FT2}
\begin{equation}
\label{hp}
H_\Lambda^{(p)}(S,J)=
-\sum_{i_1,\cdots, i_{p}\in\Lambda }
J_{i_1\cdots i_{p}}
S_{i_1}\cdots S_{i_{p}}
\end{equation}
where the couplings $J_{i_1\cdots i_{p}}$ are i.i.d. Gaussian
variables with zero average and variance 
\begin{eqnarray}
\label{wij}
E(J_{i_1\cdots i_{p}}^2)=
\frac {1}{ 2} \gamma^{(p-1)d}
\sum_{k\in \Lambda}\psi(\gamma|i_1-k|)\cdots
\psi(\gamma|i_p-k|)
\end{eqnarray}
where $\psi(|x|)$, $x\in { R}^d$, is a non-negative integrable
function verifying the normalization $\int d^dx\, \psi(|x|)=1$.  With
this choice, only variables that are at distances
$r_{ij}\; {<\atop\sim}\; \gamma^{-1}$ effectively interact.  The
effective interaction range $\gamma^{-1}$ will be assumed to be large
throughout the paper.  For technical reasons that  will be discussed later, it
is convenient to consider, instead of the pure $p$ spin Hamiltonian, a
compound Hamiltonian which mixes a $p$-spin part with $p\geq 3$ with a
(small) $p=2$ part, namely
\begin{eqnarray}
H_\Lambda(S,J)=\sqrt{a}H_\Lambda^{(2)}(S,J)+H_\Lambda^{(p)}(S,J).
\label{hami}
\end{eqnarray} 
A useful notation in the following will be 
\begin{eqnarray}
f(q)=1/2(a q^2 +q^p). 
\end{eqnarray}

Our interest for the model derives from the fact that it has been proved
that (at least for even $p$), in the thermodynamic limit, for small
$\gamma$, free-energy and local order parameter are close to the values
obtained in the corresponding long range model \cite{FT1,FT2}.  In
addition, it has been shown that the application of the replica method
to the study of effective potential functional of Ising Kac $p$ spin
for even $p$ has been shown to give rise to free-energy lower bounds
for small $\gamma$ \cite{FT3}.

The task of the next section is to set up a formalism allowing the
computation of the free-energy as a function of the overlap profile in
space $\pt(x)$ on a coarse grained scale. We consider both the
annealed case, where analogously to the space-less mean field case the
two replicas are considered on the same foot, and the quenched case,
where one of the replicas is fixed in an equilibrium reference
configuration.

\section{Replica analysis}

Similarly to the mean-field case, one can study theoretically the
annealed and the quenched potentials considering a system of
$n'=r\times n$ replicas according to the procedure explained in the
previous section.

The replica analysis of the Ising analogous of model (\ref{hp}) has
been performed in \cite{FT3}. As the local spherical case only requires minor
modifications, the derivation will only be sketched here. One has to
introduce a coarse graining length $\delta/\gamma$ such that
$1<<l<<\delta/\gamma<<1/\gamma$, and partition $\Lambda$ into boxes
$C_x$ of that size. One then considers the ``block overlap'' order
parameter matrix over these boxes
$\bQ_{\alpha,\beta}(x)=(\gamma/\delta)^d\sum_{i\in C_x} S_i^\alpha
S_i^\beta$ and rescales space by a factor $\gamma$. As in the previous
section one can write the matrix $\bQ(x)$ as a collection of $r\times r$ 
$n\times n$ matrices $Q_{a,b}^{u,v}(x)$ ($a,b=1,...,n$, and
$u,v=1,...,r$). The free-energy for fixed overlap profile $\pt(x)$ is
then obtained constraining the elements $Q_{a,a}^{1,v}(x)$ to
$\pt(x)$. 

As explained in \cite{FT3}, up to terms vanishing for small $\gamma$, 
the replica computation of the free-energy leads  to 
a coarse grained free-energy functional
\begin{equation}
F[\bQ]=\frac{1}{ \gamma^d}S[\bQ] =\frac{1}{\gamma^d} 
\int 
dx\; \left[ K( \{ { \bQ}_{\alpha,\beta}\},x)
 +V(\bQ(x)) \right]
\end{equation}
with
\begin{eqnarray}
K(  {\bQ}_{\alpha,\beta},x)&=& \frac{-\beta}{2r n}\sum_{\alpha,\beta} 
[f ( {\hat \bQ}_{\alpha,\beta}(x) )-
f({\bQ}_{\alpha,\beta}(x))]\nonumber\\
V(\bQ)&=&-\frac{1}{r n}
\left[
\frac{\beta}{2} \sum_{\alpha,\beta} f({
  \bQ}_{\alpha,\beta})+\frac{1}{2 \beta}{\rm Tr} \log \bQ 
\right]
\end{eqnarray}
and we have defined 
\begin{equation}
{\hat \bQ}_{\alpha,\beta}(x)=\int dy\; \psi(x-y)  \bQ_{\alpha,\beta}(y).
\end{equation}
The replicated partition function  $E(Z^{n'})$, could in principle be
evaluated as the functional integral 
\begin{eqnarray}
E(Z^{n'})=\int {\cal D} \bQ(x) e^{-\frac{\beta r n }{ \gamma^d}S[\bQ]}
\label{zzz}
\end{eqnarray}
where the integration only concerns the unconstrained elements of the 
matrix order parameter. 

We see that we are in presence of a replica field theory, with an action
$S[\bQ]$ that can be written as the space integral of a potential part
$V(\bQ)$, identical to the mean-field free-energy as a function of the overlap
matrix \cite{cris-som}, plus a kinetic part $K( {\bQ}_{\alpha,\beta},x)$,
sensitive to space variations of the order parameter. As stated above the
free-energy as a function of $\bQ$ is written in a closed form.

Assuming smooth variations of $\bQ$ in space, as it is customary in
non-disordered cases, one can replace the kinetic term
in the action by the lowest non trivial term in its gradient
expansion, namely
\begin{equation}
K(  {\bQ}_{\alpha,\beta},x)\approx \frac{c \beta}
{4r n }\sum_{\alpha,\beta} f''(\bQ_{\alpha,\beta})(\nabla \bQ_{\alpha,\beta})^2
\label{kin}
\end{equation}
where $c$ is the only parameter depending on the shape of the $\psi$ function
and is given by $c=\int dx\; \psi(x) x^2$. In what follows we use expression 
(\ref{kin}) for the kinetic term. With this approximation the action
$S$ becomes the space integral of a Lagrangian density, with ``coordinate
dependent'' masses $f''(\bQ_{\alpha,\beta})$. For a pure $p$-spin interaction
the mass would vanish at vanishing values of the overlaps. This is the reason
why we introduced an additional term with couple interactions in the
Hamiltonian (\ref{hami}). Without this term it would be necessary to pursue
the gradient expansion to higher orders to avoid singularities for small values
of the overlaps.\footnote{We thank J.-P. Bouchaud for this observation.}

Thanks to the large factor $\gamma^{-d}$ in front of the action,  one can
estimate the integral (\ref{zzz}) through saddle point evaluation.  One needs
then in principle to find solutions of the Euler-Lagrange equations for the
unconstrained elements of the matrix $\bQ(x)$. These take the form
\begin{eqnarray}
\frac{c\beta}{2}\nabla\cdot[f''(\bQ_{\alpha,\beta})\nabla
\bQ_{\alpha,\beta}]&=&\frac{c\beta}{4}f'''(\bQ_{\alpha,\beta})(\nabla
\bQ_{\alpha,\beta})^2-\frac{\beta}{2}f'(\bQ_{\alpha,\beta})
\nonumber\\
&&-\frac{1}{2\beta} (\bQ^{-1})_{\alpha,\beta}.
\label{field}
\end{eqnarray}
Solving these equation for fixed values of $Q_{aa}^{1,u}(x)=\pt(x)$ in space
and with the mentioned Ansatz for the remaining elements of $\bQ(x)$, leads to
expressions of the free-energy that can be continued analytically as required
in the two cases. It was proven in \cite{FT3} that the resulting expression in
the annealed case, gives at least an upper bound to the exact free-energy
functional up to terms that scale as $(L\gamma)^d$. We believe that as it
happens in the unconstrained case the expression becomes exact in the Kac
limit \cite{FT1,FT2}.

We have now an expression that allows in principle the evaluation of the
effective potential functionals. The physical free-energy is associated to the
least action space homogeneous saddle point of these functionals with respect
to $\pt(x)$. These are obtained just solving eq.s (\ref{field}) also with
respect to the elements $Q_{aa}^{1,u}(x)$. We are interested in the
regions of temperature mentioned in section 2, where the minimization of the
action admits two homogeneous solutions with degenerate or non-degenerate
minima. In the non-degenerate case we consider spherically
symmetric instantons allowing to define a free-energy barrier for
nucleation. In the degenerate case, we consider instantons with
planar symmetry in order to define an interface tension.

In the following we  denote respectively $V_a$ and $V_q$ and $K_a$ and
$K_q$ the potential and kinetic part of the action $S_a$ or $S_q$ relative to
the annealed or the quenched problem.

\section{Simple Inhomogeneous solutions}

In this section we enter in the core of our analysis and study inhomogeneous
solutions of the field equations (\ref{field}). In both annealed and quenched
cases we distinguish the region of temperatures $T_K<T<T_s$, ($T_s=T_2$ in the
annealed case and $T_s=T_c$ in the quenched case), which we call region I,
where the secondary minimum is higher then the primary one, and then
temperatures $T<T_K$, which we call region II, for which both minima are
degenerate.

Region I is where one expects  ``activated processes'' to exist
destabilizing the high free-energy minima. These kind of processes can be
studied considering inhomogeneous solution with finite action difference with
the metastable phase and unstable modes.  One has to consider then instanton
configuration such that for $r=|x|\to\infty$, the system is described by the
metastable state. The simplest of such solutions are spherical
droplets. Close to $T_K$ where the minima are nearly degenerate, one can use a
``thin wall approximation'' to derive an effective droplet model
\cite{L1,coleman}, and estimate the droplet radius and action through the
competition between a bulk free-energy gain and a surface tension loss. It has
been remarked that such approximation has to break down close to the spinodal
point $T=T_s$, where one finds ``ramified droplets'' with interface thickness
comparable to the radius. In that region the leading behavior can be obtained
by dimensional analysis of various degree of refinement \cite{K1,K2,muratov}.

In Region II the minima are degenerate and one is interested to study the
``interface between the two phases'', that one can suppose to be flat. In this
case one has to consider one-dimensional instantons that connect regions at
$\pm \infty$ with overlaps fixed at the two minima values.  The free-energy
cost for the interface is related to the value of the lower critical
dimension.

%% We seek for ``instantonic solutions'': a stationary
%% and finite action solution, connecting the secondary minimum to the
%% low overlap region.  We will limit ourselves to ``flat interfaces''
%% where the space variations occur in a definite direction ``$x$'' that
%% we shell call time. As in more usual case, the problem become then
%% formally equivalent to a mechanical problem admitting a conserved
%% quantity
%% \begin{eqnarray}
%% E=K-V
%% \end{eqnarray} 
%% that we shell call energy in the following.  

\subsection{The annealed case}

\subsubsection{Region I}

In region I the primary minimum is described by a solution where all the
overlap parameters are equal to zero, as matrices, $P=0$ and $Q=I$ this is the
solution of the unconstrained problem, and the free-energy is just equal to
twice the unconstrained free-energy.  The secondary minimum which is described
by a 1RSB solution with $q=p=\pt=\pt^*$, $q_0=p_0=0$. The value $m=m^*/2$ of
breaking parameter in the secondary minimum starts from $m^*=2$ at $T_2$ and
decreasing monotonically for decreasing temperatures, until $m^*=1$ at $T_K$.

For values of $\pt$ different from the values $0$ and $\pt^*$ that it
takes in the two minima respectively, the potential is described by
the maximum over $q$, $p$ and $m$ of the 1RSB form
\begin{eqnarray}
V_a(\pt,p,q,m)=&& -\frac{\beta}{2} [f(\pt)-(1-m)(f(q)+f(p))]\nonumber\\
&& -\frac{1}{2\beta}(1-\frac{1}{m})[\log(1-\pt -q+p)+\log(1+\pt -q-p)]
\nonumber\\
&& -\frac{1}{4\beta m}[\log(1-\pt -(1-m)(q-p))
\nonumber\\
&&
+
\log(1+\pt -(1-m)(q+p))]
\end{eqnarray}

We would like to construct a spherically symmetric solution which has
finite difference in action with respect to the uniform metastable
solution, and that takes the values of the parameters in this solution as 
boundary values for the radial distance $r$ going to infinity. 

We confine ourselves to 1RBS solutions with $m$ constant in space, and
given by the value $m^*/2$ it takes in the secondary minimum. This
choice is motivated by the fact that the breaking parameter $m$
appears as a parameter in the distribution of the free-energies of the
different ergodic components, and should not depend on space. In
addition, 1RSB matrices with values of $m$ depending on space would
not form a closed algebra and could not verify the field equations.
Within this 1RSB Ansatz we will only consider solutions such that
$q_0=p_0=0$. This is consistent with the fact that in the mean-field
model different states have vanishing overlap. We need therefore to
find a solution of the space dependent equations departing from the
secondary minimum with vanishingly small velocities for the parameters
$\pt,p$ and $q$. Unfortunately, for $T$ smaller but close to $T_2$ one
can see that for $m$ fixed to the value $m^*/2$, the point
$q=p=\pt=\pt^*$ is a maximum and not a local minimum of the function
$V_a(\pt,p,q,m^*)$ so that this Ansatz simply gives a vanishing
surface tension. While this fact could signal some inconsistency of
the present approach, it is not clear to us at present how to avoid
the hypothesis that $m=m^*/2$.  Below a second temperature $T_2'$
however, the point $q=p=\pt=\pt^*$ becomes a minimum of
$V_a(\pt,p,q,m^*)$ and we can use that Ansatz to estimate the surface
tension.

The simplest inhomogeneous solution we can look for, being described by
the metastable saddle point at infinity and having a finite
action, is described by $\pt=p=q$ in all points $x$ in space.  

With this choice, the potential part of the action becomes 
\begin{eqnarray}
&&W_a(\pt)=V_a(\pt,\pt,\pt,m^*)=\nonumber\\&&
\frac{\beta}{2}  ( 1 -  m^* )  f(\pt ) + \frac{1}{2 \beta m^*}
\left[    \left( 1 - m^* \right)  \log (1 - \pt ) - \log (1 - \pt  +  \pt  m^*)
    \right]
\end{eqnarray}

If we make the thin wall approximation\cite{L1,coleman} close to $T_K$,  
the problem becomes a one dimensional mechanical problem 
with a kinetic energy term which reduces simply to
\begin{eqnarray}
K_a=\frac {\beta c}{4} ( m^*-1) f''(\pt) \left( \frac{d\pt}{dr}\right)^2 
\end{eqnarray}
where we remember that $m^*$ takes values in the interval $[1,2]$ and 
$(m^*-1)>0$ in the region we are considering. 
 
Considering $\pt_1$ the value of $\pt<\pt^*$ such that
$W_a(\pt_1)=W_a(\pt^*)=W^*$, the 1D integral of the action density along
the instanton direction defines the surface tension $\sigma$, which
through elementary mechanics is seen to be equal to:
\begin{eqnarray}
\sigma=
\int_{\pt_1}^{\pt^*} d\pt \sqrt{2 c\beta (m^* -1)f''(\pt)[W_a(\pt)-W_a^*]}
\label{sigma}
\end{eqnarray} 
It is easy to see that close to $T_K$ the value of $\sigma$ vanishes
linearly in $T-T_K$. This is due to the fact that quite generically
 in the parameters that define the Hamiltonian, 
one has $(m^*-1)\sim W_a(\pt)-W_a^* \sim T-T_K$ uniformly $\pt$.
This estimate of the surface tension can then be exploited in a
standard way to complete the calculation and compute the radius and
free-energy of the critical droplet. 

In dimension $d$  the action of a droplet with radius $R$ is given by
\begin{eqnarray}
\Delta S_a(R)= -\Delta V_a v_d R^d+\sigma s_d R^{d-1}
\label{drop}
\end{eqnarray}
where we have denoted as $s_d$ and $v_d=s_d/d$ respectively the
unitary spherical surface and volume. $\Delta V_a $ is the
free-energy difference between the secondary and the primary minimum,
which, as already noted, close to $T_K$ scales like $\Delta V\approx
(T-T_K)^2$. Maximization with respect to $R$, gives the radius $R^*$
of the critical droplet where the free-energy (\ref{drop}) is maximum:
$R^*=\frac{(d-1)\sigma}{\Delta V}$. Given the scaling of $\Delta V$
and $\sigma$, one sees that close to $T_K$ it behaves as $R^*\approx
(T-T_K)^{-1}$ leading for the free-energy of the critical droplet
\begin{eqnarray}
\Delta F^*\approx \frac{const}{(T-T_K)^{d-2}}. 
\end{eqnarray}
This formula has the attractive feature of having the structure of the
Vogel-Fulcher law in three dimensions, and, despite a finite surface
tension in region I, would indicate a lower critical dimension for
1RSB transitions equal to 2. Notice that this divergence of the
free-energy stems from the fact that both the difference of
free-energy between the two minima and the surface tension vanish with
different exponents at the transition. Unfortunately, it is not clear
to us at present how and if $\Delta F^*$ could be related to a
free-energy barrier governing the relaxation time of the liquid phase.
One would expect $\Delta F^*$ to be the free-energy barrier to
equilibration for a system prepared in one of the internal free-energy
ground states.  The equilibrium relaxation time should rather be
associated with the quenched potential barrier, and being shorter than
the one necessary to relax an internal free-energy ground state. We
will see that in that more physical case our solution 
surprisingly gives a higher free-energy barrier.

In figure \ref{pot} we show the result of formula (\ref{sigma}) as a function
of temperature in the whole temperate range $T_K<T<T_2$. It should be
remembered (see below) that the flat interface approximation breaks
down at $T_2'$. Analogously, the space
dependence of the instantonic solution $\pt(x)$ can easily be
obtained, and its typical shape is displayed in fig. \ref{ist}. The
typical spatial extension of the instanton is given by 
\begin{eqnarray}
\xi=\left( \frac{1}{\beta c (m^*-1)f''(\pt^*)}
\frac{d^2 W(\pt^*)}{d \pt^2}\right)^{-1/2}
\end{eqnarray}
As in usual cases this diverges as $\xi\sim |T-T_2'|^{-1/2}$ close to
the ``spinodal temperature'' $T_2'$ and the thin walls approximation
breaks down in $D>1$. This is at variance with what would be the typical
spinodal behavior $\xi\sim |T_2'-T|^{-1/4}$. In fact the reason for
this anomaly is the mentioned fact that the point $T_2'$ is the point
where the true minimum of the potential passes from being a maximum
to a minimum when the parameter $m$ has been fixed
to $m^*$. As a result, around $\pt_t=\pt^*|_{T=T_2'}$, the potential
can be approximated as $W(\pt)\approx A+ a (\pt
-\pt_t)^3-b(T-T'_2)^2(\pt -\pt_t)$, with $A$, $a$ and $b$ smoothly
varying functions of $T$. This gives rise to a free-energy barrier
scaling as $|T-T_2'|^{3-d/2}$, which, 
while giving an upper critical
dimension equal to 6, differs from the usual spinodal scaling with an
exponent $3/2-d/4$ \cite{K1,K2}. As in non-disordered systems we
expect that higher order in the expansion of the free-energy
functional should be taken into account above dimension 6
\cite{muratov}.  
\begin{figure}
\begin{center}
\epsfxsize=350pt
\epsffile{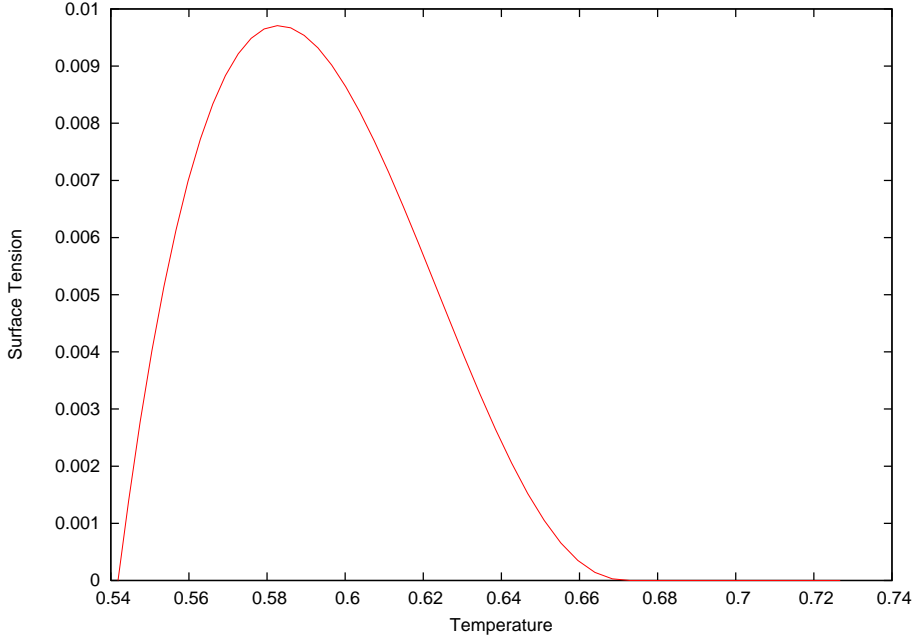} 
\end{center}
  \caption[0]{\protect\label{sig}  
1D instanton action in region I in the annealed case. We considered
 $p=4$ and $a=0.1$. In this case the critical temperatures are:
 $T_2=0.727$ $T_2'=0.673$ and $T_K=0.542$. The surface
 tension vanishes both at $T_2'$ and $T_K$. The surface tension
 vanishes linearly at $T_K$.}
\end{figure}

%% \begin{figure}
%% \begin{center}
%% \epsfxsize=350pt
%% \epsffile{droplet.eps} 
%% \end{center}
%%   \caption[0]{\protect\label{dro}  
%% Radius of the critical droplet as a function of temperature for
%% the same parameters as in fig. \ref{sig} }
%% \end{figure}
\begin{figure}
\begin{center}
\epsfxsize=350pt
\epsffile{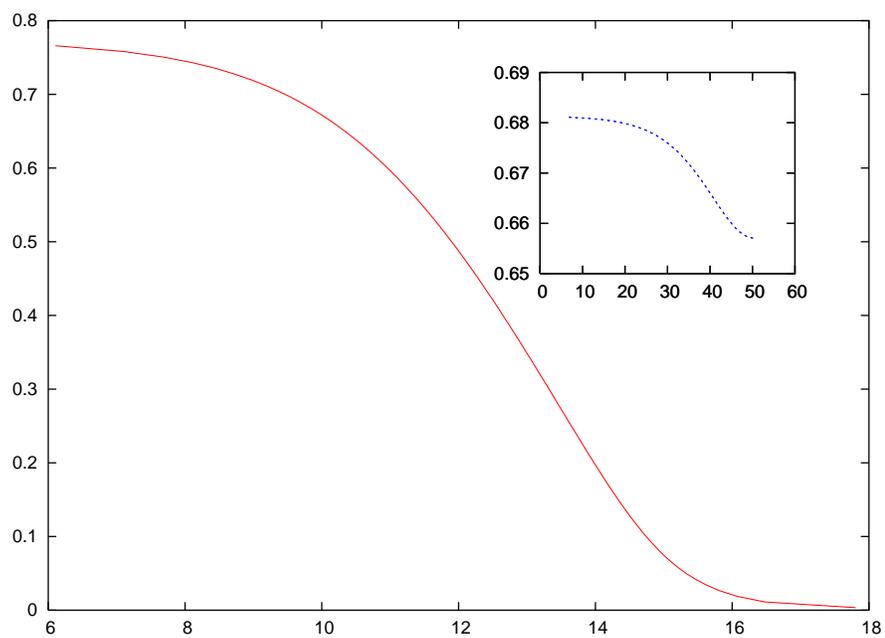} 
\end{center}
  \caption[0]{\protect\label{ist}  
Shape of the instanton for the same parameters ad fig. \ref{sig} and
$T=T_K=0.542$. At this temperature the annealed and quenched instanton
coincide. In the inset the annealed instanton for $T=0.667$. }
\end{figure}

\subsubsection{Region II}

Let us now consider briefly the region II.
 Below $T_K$ both the
primary minimum and the secondary minimum are described by 1RSB
solutions.  The primary minimum  has $\pt=0$,
$q(u)=q_{EA}\theta(u-m^*)$ and $p(u)=0$. The secondary minimum has
$\pt=q_{EA}$ and $q(u)=p(u)=q_{EA}\theta(u-m^*/2)$. $q_{EA}$ and $m^*$
are respectively the value of the overlap and the breaking parameter
that appear in the 1RSB solution of the unconstrained problem. In order to
interpolate in space between these two solutions it is natural to
consider a two step RSB form of the kind
\begin{eqnarray}
&&q(u)=\left\{ 
\begin{array}{cc}
0 & 0\leq u< m^*/2\\
q_0 & m^*/2\leq u< m^*\\
q_{1} & u\geq m^*
\end{array}
\right.\nonumber\\
&&
p(u)=\left\{ 
\begin{array}{cc}
0 & 0\leq u< m^*/2\\
p_0 & m^*/2\leq u< m^*\\
p_{1} & u\geq m^*
\end{array}
\right.
\end{eqnarray}
where $q0,q_1,p_0,p_1$ and $\pt$ depend on space, while $m^*$ is fixed.
Correspondingly, the potential takes the form:
\begin{eqnarray}
&&V_a(q_0,q_1,p_0,p_1,\pt)= -\frac{\beta}{2}
\left[ -(1-m^*)f(q_1)-\frac{m^*}{2}
  f(q_0)+
\right.
\nonumber\\
&&\left.f(\pt)-(1-m^*)f(p_1)-\frac{m^*}{2} f(p_0)\right]
\nonumber\\
&&-\frac{1}{2\beta m^*} \left\{
\log[(1-(1-m^*)q_1-\frac{m^*}{2} q_0)^2-(\pt-(1-m^*)p_1-\frac{m^*}{2}
p_0)^2]
\right.\nonumber\\
&&-\frac{1}{2} 
\log[(1-(1-m^*)q_1-m^* q_0)^2-(\pt-(1-m^*)p_1-m^*
p_0)^2]
+
\nonumber\\
&&\left.\frac{1}{2}(m^*-1)
\log[(1-q_1)^2-(\pt-p_1)^2]\right\}
\end{eqnarray}
while the kinetic part of the action is:
\begin{eqnarray}
&&K_a(q_0,q_1,p_0,p_1,\pt)= \frac{c\beta}{2}
\left[ -(1-m^*)f''(q_1)(\nabla q_1)^2-\frac{m^*}{2}
  f''(q_0)(\nabla q_0)^2+
\right.
\nonumber\\
&&\left.f''(\pt)(\nabla \pt)^2-(1-m^*)f''(p_1)(\nabla
  p_1)^2-\frac{m^*}{2} 
f''(p_0)(\nabla q_0)^2\right]
\end{eqnarray}
Simple differentiation reveals the existence of a solution such that
in all points in space $q_0=p_0=p_1=\pt$, while $q_1$ is independent
of position and equal to the value $q_{EA}$ of the Edwards-Anderson
parameter of the model in the low temperature phase. Assuming a flat
interface with space variations in the direction $x$, the equation
verified by $\pt$ is: 
\begin{eqnarray}
&&c\beta f''(\pt) \frac{\partial^2 \pt}{\partial x^2}=
\nonumber\\
&&c\beta f'''(\pt) \left(\frac{\partial \pt}{\partial x}\right)^2-\beta f'(\pt)
\nonumber\\
&&-\frac{1}{2\beta m}
\left\{
\frac{1}{1-(1-m)q_{EA}-m \pt}
-\frac{1}{1-(1-m)q_{EA}}\right\}. 
\label{oops}
\end{eqnarray}
This equation has a solution that corresponds to a well defined
instanton in space, continuing smoothly the form found in the region I
in the thin wall approximation. Interestingly enough, one readily
verifies that the various terms compensate each other in such a way
that the solution has an identically vanishing interface cost. In the
replica formalism this is related to the fact that the number of
matrix elements equal to $\pt(x)$ is proportional to $n$ and is
reminiscent of a similar result of \cite{FPV3} for the
Edwards-Anderson model, where it was found a replica matrix with vanishing
excess action. The important difference though is that here the matrix
with vanishing excess action is a true solution of the saddle point
equations. At present we do not have a clear interpretation of this
result. Probably the computation of the determinant of the small
fluctuation matrix around the solution would give more informations
about its nature.  It is also possible that the vanishing of the
surface tension is a signal that another solution to the
Euler-Lagrange equations exists, that has an action scaling as
$L^{d-d_c}$ with $d_c>1$. Unfortunately for the time being we did not
find such a solution. We will see in the next paragraph as the
quenched formalism gives rise to the same instanton, but this time
with a finite action.

\subsection{The quenched case} 

Though with a different physical meaning, the analysis of this case
parallels closely the one for the annealed case. If we
interpret the instanton action as a free-energy barrier, in this case
we directly relate to the equilibrium relaxation time of the system.
The computation of the critical droplet in the quenched case, will
allow a fundamental computation of the glass coherence length as
proposed by Biroli and Bouchaud.

The most general expression for the potential that we use is a
1RSB form with $T=Q$ and $S=P$. One can check that is is always a 
solution to the saddle point equations and that both minima of the
potential are correctly described by this form in all temperature
regions. With this Ansatz we have:
\begin{eqnarray}
V_q(\pt,p,q,m)=&&-\frac{\beta}{2} [f(\pt)-(1-m)(f(q)+f(p))]
\nonumber\\
&&
-\frac{1}{2\beta}
\left\{
\frac{1}{m} \log(1-\pt-(1-m)(q-p))
\right.
\nonumber\\
&&
\left.
+(1-\frac{1}{m}) \log(1-\pt-q+p)
\right.
\nonumber\\
&&
\left.
\frac{1}{m} \frac{(\pt-(1-m)p)}{1-(1-m)q}
\right.
\nonumber\\
&&
\left.
+(1-\frac{1}{m}) \frac{\pt-p}{(1-q)}
\right\}
\end{eqnarray}
\begin{eqnarray}
K_q(\pt,p,q,m)=\frac{c\beta}{2} [f''(\pt)(\nabla
\pt)^2-(1-m)(f''(q)(\nabla q)^2+f''(p)(\nabla p)^2)]
\end{eqnarray}

We start from the analysis of the minima of the potential as a
function of the overlap. The quenched potential develops the secondary
minimum at $T_c$, the temperature of the dynamical transition. In the
temperature range $T_K<T<T_c$, that defines region I, the minima have
the following structure: the primary minimum is described by $Q=T=I$,
$P=S=0$, the secondary minimum has the same $Q$ and $T$, while $P$ and
$S$ have the diagonal form $P=S=\pt^* I$.

At low temperature, in the region II $T<T_K$ on the other hand both
the primary and the secondary minima are described by 1RSB solutions.
In the primary minimum, one has that $Q=T$ and both matrices are
parameterized by the function $q(u)$ which takes the 1RSB form $q(u)=q_{EA}
\theta(u-m^*)$, while $P=S=0$. The secondary minimum, has the same $Q$
and $T$, while $P=S$ are parameterized by a diagonal element $\pt=q_{EA}$
and the function $p(u)=q_{EA} \theta(u-m^*)$. For both minima the values of
$m^*$ and $q_{EA}$ are the same and coincide with their equilibrium values
in the unconstrained free-energy.  In both regions we now seek 
solutions that connect the two minima.  

\subsubsection{Region I}

Since in this case both the primary and the secondary minima are described a
replica symmetric saddle points, we choose a RS solution to describe the
instanton. It is easy to see that the matrix $Q$ verifies equations that are
uncoupled from the remaining matrices and coincide with the ones for the
unconstrained system. This is coherent with the fact that the reference state
used in the quenched potential is unaffected by the coupling. One should then
choose for $Q$ the space independent solution which is appropriate for the
temperature range at hand.  In this region one has $Q=I$.  A simple inspection
shows that the equations admit solutions with $T=Q$ and $S=P$ in all points of
space, and this will be our choice, with with $T=Q=I$ and $P=S=\pt I$.  The
typical shape of the surface tension in the thin wall approximation as a
function of the temperature is given in figure \ref{sig-a}.  We see that in
this case the surface tension stays finite at $T_K$.  The computation of the
radius and free-energy of the critical droplet, follows the same route as in
the annealed case.  Taking into account that the difference between the two
minima is the configurational entropy, which vanishes linearly as $T-T_K$, at
$T_K$, one finds that the radius of the critical droplet is given by $R^*\sim
\sigma/(T\Sigma)\sim (T-T_K)^{-1}$, while its free-energy is $\Delta
F^*\approx (T-T_K)^{-(d-1)}$ as in conventional first order transitions.  It
is interesting to compare in detail the behavior of the various quantities
close to $T_K$ with the corresponding one in the annealed case.  To the
leading order in $T-T_K$ one finds $\Delta V_a=(m^*-1) \Delta V_q$,
$W_a=(m^*-1) W_q$. Consequently, $\sigma_a=(m^*-1)\sigma_q$ and $\Delta
F_a=(m^*-1) \Delta F_q$, one verifies also that remarkably, despite a
different surface tension and bulk free-energy gain $\Delta V$, the
characteristics of the droplets, instanton shape and radius are the same in
the two approaches. We notice that given the difference in $\Delta V_a$ and
$\Delta V_q$ the behavior we find for $\sigma_a$ and $\sigma_q$ is the only
one compatible with the fact of having the same droplet describing the two
cases.

Close to $T_c$, one finds in the quenched case the conventional mean-field
spinodal scaling, where the instanton has a width $\xi\sim |T-T_c|^{-1/4}$
while the free-energy barrier scales like $\Delta F^* \sim
|T-T_c|^{3/2-d/4}$. One indeed can verify that all this holds because in this
case one has the usual form of the potential close to the high minimum:
$W(\pt)\approx A+ a (\pt -\pt_t)^3-b(T-T_c)(\pt -\pt_t)$.

\subsubsection{Region II}

\begin{figure}
\begin{center}
\epsfxsize=350pt
\epsffile{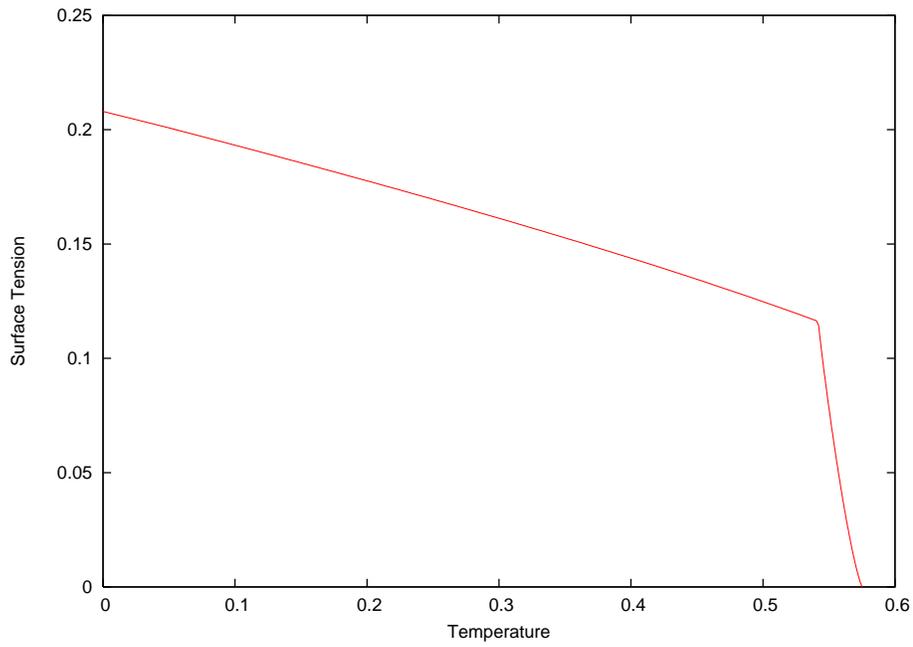} 
\end{center}
  \caption[0]{\protect\label{sig-a}  
1D instanton's action 
 in regions I and II in the quenched case. We considered
 $p=4$ and $a=0.1$. In this case the critical temperatures are:
 $T_c=0.575$ and $T_K=0.542$. In this case the instanton has non
 vanishing action in both regions.} 
\end{figure}

In the low temperature phase, region II, the solution for the matrix $Q$ is of
the 1RSB type with parameters $m^*$ and $q=q_{EA}$ independent of
position. Again we choose the solution which has $T=Q$ independent of position
and $S=P$. The 1RSB solution for $P$ will be parameterized by the value of the
breaking parameter, which naturally will be taken equal to $m^*$ and the
values of the space dependent diagonal overlap $\pt$ and off-diagonal overlap
$p$. Once again, if we choose $p=\pt$ the problem becomes one dimensional and
can be trivially integrated.  The field equation for $\pt$ and thus the
resulting instanton solution coincide with eq.  (\ref{oops}) found in the
annealed case in all region II. Despite that, thanks to the different kind of
analytic continuation, the value of the surface tension is different from zero
in all region II, and connects continuously with the value found in region I
at $T_K$.  The typical behavior of the quenched surface tension is seen in
figure \ref{sig-a} and indicates a critical dimension $d_c=1$. In order to
investigate the possibility that this result is an artifact of our simple
solution, we have looked for a different 1RSB saddle point,
$t(u)=s(u)=\tilde{s} \theta(u-m)$ and $p(u)=\pt \theta(u-m)$. Unfortunately,
this solution, while giving rise to different values of $\sigma$ for $T<T_K$,
does not affect the scaling of the surface tension.

\section{Summary and Conclusions}

In this paper we set up a general formalism to study nucleation and
phase coexistence in terms of inhomogeneous solution of field
theoretical equations for disordered models with long but finite
interaction range. We presented here the simplest instantonic solutions
to the replica saddle point equations for a $p$-spin model with a
local spherical constraint. We used two different kinds of set-ups: the
annealed potential and the quenched potential, for which we have
critical droplet to nucleation in region I and flat coexistence
interfaces in region II.

As in usual nucleation theory, in region I we have found -in the thin wall
approximation- spherically symmetric solutions that leave the unstable minimum
of the potential function at infinity with vanishing velocity. Differently
from cases with scalar order parameter where at most one solution with a given
spatial symmetry can exist, here multiple solutions are in principle
possible. Indeed in the replica formalism one has to choose an Ansatz to
parameterize the overlap matrix, and different parameterizations could lead to
different solutions. In this paper, we have always performed the simplest
choice.  In the case of the annealed potential we find a solution such that
the surface tension has non zero value in a temperature range $T_K<T<T_2'$
while is zero below the static transition temperature $T_K$. Its linear
behavior close to $T_K$ implies a critical droplet with free-energy scaling as
$(T-T_K)^{-(d-2)}$. In the quenched case we find different results: the
surface tension is different from zero for all temperatures below the dynamic
transition temperature $T_c$. It has a finite value at $T_K$ where there is a
discontinuity in its derivative.  Correspondingly, the free-energy of the
critical droplet scales as $(T-T_K)^{-(d-1)}$.  In region II one finds that
the same instantonic solution describes the quenched and annealed
interface. This is coherent with the fact that (differently from region I) in
the two different set-ups, the minima of the potential functions are
associated to the same family of states. Unfortunately, due to different
analytic continuations, one finds different values of the interface
tension. In the annealed case the interface tension is vanishing in all region
II, in the quenched case it has a finite value. This is a paradoxical
situation indicating possibly that more complicate replica solutions should be
used to describe one or both situations. Further research and a deeper
analysis of the field equations are needed to clarify the issue.

There are some obvious issues that we have not addressed in this paper.  The
first one concerns the physical interpretation of space-dependent
overlap matrices. In Mean Field, overlap matrices are interpreted as
describing global correlation functions \cite{mpv}. It is of course
tempting to interpret local matrices as generators of local
correlation. We did not attempt in this paper an analysis along that
lines. A detailed study should be devised to substantiate
this hypothesis. 

The second issue concerns the relation between the free-energy barriers
derived within the present approach and the relaxation time to equilibrium.
In Ising like transitions, one can model the transition kinetic by
time-dependent Landau-Ginzburg equations with noise and in this way relate the
nucleation barrier to the rate of relaxation of metastable phases
\cite{L1,L3}.  This kind of approach has been rigorously shown to represent
the spinodal decomposition dynamics in the scaling regime for Kac kind
interactions \cite{presutti}. In the disordered case it is at present not
clear if one could write the time evolution of the space-dependent overlap as
a gradient equation in the free-energy functionals defined in this
paper. Consequently, one can not say which, if any, between the annealed or
the quenched droplet free-energy can be associated to the main relaxation time
of the system. A proper theory of relaxation rates should start from a
dynamical approach similar to the the mean-field one, but in which the
dynamical order parameters are allowed to depend on space.  Again the use of
Kac models should allow to obtain the theory as an expansion around Mean
Field.

Summarizing, this paper indicates a route to study 
ergodicity restoration and the possibility of ideal glassy phases in
finite dimension. Further studies are necessary to overcome the
many obstacles found in the pathway. 

\section*{ Acknowledgments}

We thank G. Bianconi for interactions at an initial stage of this work,
G. Biroli for many useful exchanges and careful reading of the manuscript and
G. Parisi for an important discussion. This work was supported in part by the
European Community's Human Potential programme under contract
``HPRN-CT-2002-00319 STIPCO''.

\section{Appendix}

In this appendix we derive some of the formulae used in our
computations.

The form for the $rn\times rn$ replica matrix $\bQ$ discussed in
section 2 reads

\begin{eqnarray}
\bQ=\left(
\begin{array}{cccccc}
Q&P&P&P&P&...\\
P&T&S&S&S&...\\
P&S&T&S&S&...\\
P&S&S&T&S&...\\
P&S&S&S&T&...\\
...&...&...&...&...&...
\end{array}
\right)
\end{eqnarray}
where each symbol represents an $n\times n$ matrix. 

Using this form one readily finds for the potential part of the free-energy: 

\begin{eqnarray}
&&V(\bQ)=-\frac{\beta}{2}\sum_{a,b}^{1,n}\left[ f(Q_{a,b})+2(r-1)f(P_{a,b})+(r-1) 
f(T_{a,b})+(r-1)(r-2)f(S_{a,b})\right]\nonumber\\
&&-\frac{1}{2\beta} [(r-2){\rm Tr} \log(T-S)+{\rm Tr}
\log[Q(T+(r-2)S)-(r-1)P^2]
\end{eqnarray}
and for the kinetic one
\begin{eqnarray}
&&K(\bQ)=\frac{c\beta}{2}\sum_{a,b}^{1,n}\left[ f''(Q_{a,b})(\nabla Q_{ab})^2
+2(r-1)f''(P_{a,b})(\nabla P_{ab})^2+\right.\nonumber\\&&
\left
(r-1) 
f''(T_{a,b})(\nabla T_{ab})^2+(r-1)(r-2)f''(S_{a,b})(\nabla S_{ab})^2\right]
\end{eqnarray} 
From this form the Euler-Langrange equations for the various matrices can
easily be written.

The most general Ansatz we use is such that for all points in space the
structure of the various matrices is of the 1RSB type described in
section II. The overlap parameters are space dependent, while the
breaking point $m$ is equal for all points in space. The detailed
parameterization is such that each matrix is parameterized by a single
diagonal element, an single off-diagonal element, and a common
breaking parameter. The matrices $Q$ and $T$ have diagonal element
equal to 1 thanks to the local spherical constraint and non diagonal
elements $q$ and $t$ respectively. The matrices $P$ and $S$ have
respectively diagonal elements $\pt$ and ${\tilde s}$ and non diagonal
elements $p$ and $s$. Inserting, with self-evident notation the 1RSB
Ansatz and introducing the notation $\langle q\rangle =(1-m)q$ one has
\begin{eqnarray}
&&V=-\frac{\beta}{2}\left[ f(1)-\langle f(q)\rangle
+2(r-1)(f(\pt)-\langle f(p)\rangle)\right.
\nonumber\\&&\left. +(r-1) (f(1)-\langle f(t)\rangle)
+(r-1)(r-2)(f({\tilde s})-\langle f(s)\rangle) \right]\nonumber\\
&&-\frac{1}{2\beta} \left\{
(r-2)\left[
\frac 1 m \log(1-{\tilde s}-\langle t-s\rangle) +
(1-\frac 1 m) \log(1-{\tilde s}- (t-s))
\right]\right.
\nonumber\\&&
\left. 
\frac 1 m \log[(1-\langle q\rangle )(1-\langle t\rangle +(r-2)({\tilde s}
-\langle s\rangle))-(r-1)(\pt-\langle p\rangle)^2]
\right.
\nonumber\\&&
\left.
+(1-\frac 1 m)\log[(1- q )(1- t +(r-2)({\tilde s}- s))-(r-1)(\pt- p)^2]
\right\}
\end{eqnarray}
\begin{eqnarray}
&&K=\frac{c\beta}{2} \left\{
-\langle f''(q)(\nabla q)^2\rangle
+2(r-1)[f''(\pt)(\nabla \pt)^2-\langle f''(p)(\nabla p)^2\rangle]
\right. 
\nonumber
\\
&&
\left. 
+(r-1)[f''(\tilde{t})(\nabla \tilde{t})^2-\langle f''(t)(\nabla t)^2\rangle]
+(r-1)(r-2)[f''(\tilde{s})(\nabla \tilde{s})^2-\langle f''(s)(\nabla 
s)^2\rangle]
\right\}
\end{eqnarray}
One can specify to the annealed two replica case just setting $r=2$
and to the quenched case taking the derivative with respect to $r$ and
setting $r=1$.

\end{document}